\documentstyle[12pt,aasms4]{article}

\begin{document}

\title{The Chemical Compositions of the SRd Variable Stars-- II. 
WY\,Andromedae, VW\,Eridani, and UW\,Librae} 
\author{
Sunetra Giridhar}
\affil{Indian Institute of Astrophysics;
Bangalore,  560034 India\\
giridhar@iiap.ernet.in}

\author{David L.\ Lambert }
\affil{Department of Astronomy; University of
Texas; Austin, TX 78712-1083
\\ dll@astro.as.utexas.edu}

\author{Guillermo Gonzalez}
\affil{Department of Astronomy;
University of Washington;
Seattle, WA 98195-1580\\gonzalez@astro.washington.edu}

\begin{abstract}

Chemical compositions are derived from high-resolution spectra for three stars
classed as SRd variables in the {\it General Catalogue of Variable Stars}.
These stars are shown to be metal-poor supergiants:
WY\,And with [Fe/H] = -1.0, VW\,Eri with
[Fe/H] = -1.8, and UW\,Lib with [Fe/H] = -1.2.  
Their compositions are identical to within the measurement errors with
the compositions of subdwarfs, subgiants, and less evolved
giants  of the same [Fe/H].  The stars are
at the tip of the first giant branch or in the early stages of evolution
along the asymptotic giant branch (AGB). 
There is no convincing evidence that these SRd variables
are experiencing thermal pulsing and the third dredge-up on the AGB. 
The SRds appear to be the cool limit of the sequence of 
RV\,Tauri variables.

{\it Subject headings: stars:abundances -- stars:variables:other (SRd) --
stars:AGB and post-AGB}

\end{abstract}

\section{Introduction}

This series of papers  presents and discusses determinations of the chemical
compositions of the SRd variables  for which the {\it General Catalogue of
Variable Stars} provides the following prosaic definition:  ``Semiregular
variable giants and supergiants of spectral types F, G, K sometimes with
emission lines in their spectra.'' 
This definition admits  massive supergiants (e.g., $\rho$\,Cas)
and high-velocity metal-poor low-mass supergiants (e.g., TY\,Vir)
to the same club. Our goal is to identify and analyse the metal-poor low mass
stars by undertaking detailed abundance analyses of  SRd
variables listed in the GCVS. Presently, we have observed approximately 30
variables. Our primary interest is in the metal-poor low-mass stars.

In our first paper (Giridhar, Lambert \& Gonzalez
 1998a\markcite{Gir98a}, Paper I),
 we discussed  four stars:
XY\,Aqr, RX\,Cep, AB\,Leo, and SV\,UMa. The first two were shown to be
of approximately
solar metallicity and probably not variable stars.  Both AB\,Leo and SV\,UMa
were demonstrated to have a
low metal abundance ([Fe/H]
$\simeq -1.5$) and are certainly  variable stars. 
Here, we discuss three acknowledged variables (Table\,1)
 of high radial velocity
and show that they are indeed metal-poor supergiants.

\section{Observations and Abundance Analyses}

Spectra were obtained at the McDonald Observatory with either the 
 2.1 m telescope equipped
with a Cassegrain echelle spectrograph and a Reticon 1200 $\times  400$ pixel
CCD (McCarthy et al.\ 1993\markcite{McC93}) or the 2.7m telescope and the
{\it 2dcoud\'{e}} echelle spectrograph (Tull et al. 1995\markcite{Tul95}).
Spectra were reduced and analysed by procedures
 described in
Paper I.
Atmospheric parameters derived principally from the Fe\,{\sc i} and
Fe\,{\sc ii} lines are given in Table\,1.

Our derived abundances are quoted to $\pm 0.1$\,dex, and given in Table 3 
as [X/H] = $\log \epsilon$(X/H) $ - \log \epsilon_{\odot}$(X/H) and [X/Fe]
 where the H
abundance is on the customary scale, and solar abundances are taken from
Grevesse, Noels, \& Sauval (1996)\markcite{Gre96}. The total error in the
absolute abundance of a well observed element may be
 about $\pm 0.2$ dex when the
various sources of error (equivalent width, effective temperature, etc.) are
considered. This does not include systematic errors arising, for example,
from the adoption of LTE and the neglect of hyperfine splitting.

\section{Results and Discussion}

\subsection{VW\,Eridani}

The SRd variables that are low-mass supergiants are cool stars
with spectra that may contain TiO bands. Since these bands
contribute many lines and reduce the number of unblended lines,
we discuss the stars in order of increasing TiO band strength.
Not surprisingly the star not showing TiO bands is the most
metal-poor of the trio.

 Our analysis shows VW\,Eri
to have [Fe/H] = -1.8. 
This SRd is assigned a period of 83.4 days in the GCVS
(Kholopov et al. 1985) 
and a high radial velocity by Preston (1971\markcite{Pre71}).
 Eggen (1973\markcite{Egg73})
published UBVRI photometry. Our spectrum and analysis fully 
confirm that the star is a metal-poor supergiant. The
spectrum gives a heliocentric radial velocity of 146.5 $\pm$
0.6 km s$^{-1}$, a value in agreement with Preston's
results. 

Our derived abundances as [X/Fe] are compared in Table\,4
with those expected of a [Fe/H] = -1.8 star from analyses
of metal-poor dwarfs and giants.
Local field dwarfs and giants with very
few exceptions show a common [X/Fe] at a given [Fe/H]. It is
this common [X/Fe] for [Fe/H] = -1.8 that is given in Table\,4.
Sources for the expected [X/Fe] are as follows:
Israelian et al. (1998\markcite{Isr98}) and Boesgaard et al.
(1999\markcite{Boe99}) for O;
 Pilachowski, Sneden \& Kraft (1996\markcite{Pil96})
 for Na; Gratton \& Sneden (1988\markcite{Gra88})
for Mg; McWilliam (1997\markcite{McW97}) for Al and Eu; Gratton \& Sneden
(1991\markcite{Gra91}) for Si, Ca, Sc, Ti, V, Cr, Mn, Co, and Ni;
Gratton (1989\markcite{Gra89}) for Mn; Sneden, Gratton \& Crocker
(1991\markcite{Sne91}) for Cu and Zn; Gratton \& Sneden 
(1994\markcite{Gra94}) for Y, La, Ce, Pr, Nd, and Sm.
All of these references review previous literature on the
elemental abundances and note the generally close agreement
between the referenced results and other results. For many
elements, the expected value of [X/Fe] should be accurate to
about $\pm$ 0.1 dex. The lack of scatter in [X/Fe] at a given
[Fe/H] for samples composed of stars now in the solar neighborhood
but originating from quite different parts of the Galaxy suggests
that a SRd like VW\,Eri should have the expected pattern of 
abundances.

There is surprisingly and probably fortuitously
 good agreement between the measured and common [X/Fe].
In particular,   characteristic signatures of a metal-poor
dwarf are found in the measured [X/Fe] of VW\,Eri: notably,
 an overabundance
of the $\alpha$-elements,  an underabundance of  Mn, and the  underabundance
of Cu in the presence of a normal Zn abundance.  
Two elements from Na to Eu with a difference of greater than $\pm$ 0.3 dex 
between observed and expected abundance are  V and Eu. We assume that
these differences are simply due to above average errors of measurement.
Neglect of hyperfine splitting is not a likely soure of error because the
lines are weak. A more probable source is the possibility of blends affecting
the weak lines. The strongest V\,{\sc i} line, a 31 m\AA\ line, gives an
abundance 0.3 dex less than the next strongest line at 14 m\AA.

The oxygen abundance is based on the 6300\AA\ [O\,{\sc i}]
line and the O\,{\sc i} lines at 7774 and 7775\AA. The
forbidden line gives a systematically lower abundance by about
0.4 dex. The mean abundance corresponds to [O/Fe] = 0.8 with the
[O\,{\sc i}] line giving [O/Fe] = 0.6.  This discrepancy
between forbidden and permitted lines is similar to the
discrepancy between the same lines in spectra 
of subdwarfs. Israelian et al. (1998\markcite{Isr98})
 obtain consistent determinations
of the O abundance from OH ultraviolet  and  the
O\,{\sc i} 7770-7775\AA\ lines. 
Boesgaard et al. (1999\markcite{Boe99}) in an independent
analysis  obtain an identical result. The mean
estimate from these recent determinations
 given in Table\,4 is in good agreement with our
measurement.

%Carbon from three C\,{\sc i}
%lines is underabundant relative to iron with [C/Fe] = -0.5. 
%The initial abundance may be inferred by combining the recent
%estimates of [O/Fe] with
%the C/O ratio derived from  permitted C\,{\sc i} and O\,{\sc i}
%lines by Tomkin et al. (1992\markcite{Tom92}).
%Between [Fe/H] = -1 and -2, [C/O] = -0.4
%from Tomkin\markcite{Tom92} et al. and, hence [C/Fe] = 0.3 for the initial
%abundance.  Our measurement of [C/Fe] = -0.5 implies a stiff
%reduction in excess of predictions for the first dredge-up. 

Heavy elements are well represented in the spectrum. 
The Ba\,{\sc ii} lines are rejected as unsuitable because
they are very strong with an
indication  that they are contaminated by a 
circumstellar component. With one exception, the
heavy elements give [X/Fe] in the range -0.2 to +0.2,
as expected from Zhao \& Magain (1991\markcite{Zha91}) and
Gratton \& Sneden (1994\markcite{Gra94}), i.e., the star has not
experienced enrichment of $s$-process products through the
third dredge-up.
The apparent exception is Eu with [Eu/Fe] = 0.6. 
Observations of Eu show that this $r$-process element
is enriched in normal metal-poor stars: Gratton
 \& Sneden (1994\markcite{Gra94} --
see also McWilliam's [1997\markcite{McW97}]
compilation) find 
[Eu/Fe] $\simeq$ 0.3, a value  less than our estimate
based on 2 Eu\,{\sc ii} lines. 

In summary, VW\,Eri judged by composition is a normal red giant that
has experienced the first dredge-up but  not the third dredge-up
on the AGB.

\subsection{UW\,Librae}

This SRd with a period of 84.7 days
was studied extensively at low-dispersion by Joy (1952\markcite{Joy52})
who noted the spectral type to vary from G0 to K4 and the radial
velocity from 142 km s$^{-1}$ to 194 km s$^{-1}$. Photometry was
provided by Eggen (1977\markcite{Egg77}).
Our analysis is based on two Sandiford echelle spectra from 1995
June 21 and 23. The heliocentric radial velocity of 166 $\pm 2$
km s$^{-1}$ is within the range reported by Joy. The spectra
provide coverage from 4450 - 4940 \AA\ and 5770 - 7240 \AA. 
Bands of TiO  are quite prominent. 
The crowded spectrum limited our
selection of useful lines.  Results of the abundance analysis
are given in Table 3. The star is clearly a metal-poor
supergiant.\footnote{Dawson (1979\markcite{Daw79})
 argued on the basis of reddening-corrected
DDO photometry that UW\,Lib was a G dwarf of near-solar composition.}

Iron is  underabundant with [Fe/H] = -1.3. 
We consider UW\,Lib to
have the expected composition for its [Fe/H]. The latter may
be obtained for most elements by linear interpolation between the values
given in Table\,4 for [Fe/H] = -1.8 and [X/Fe] = 0 at [Fe/H].
The $\alpha$-elements have approximately the same overabundance
at [Fe/H] = -1.0 as at -1.8.
Agreement between observed and expected values of [X/Fe] is not
as good as in the case of VW\,Eri. We attribute this to the
presence of TiO lines in many portions of the spectrum.

 Oxygen based on the
6300 \AA\ and 6363\AA\ [O\,{\sc i}] lines has the abundance
[O/Fe] = 0.7, which is consistent with recent
measurements 
(Israelian et al. 1998\markcite{Isr98};
 Boesgaard et al. 1999\markcite{Boe99}). The
traditional $\alpha$-elements have [$\alpha$/Fe] =
0.4 (Mg), 0.0 (Ca),  and 0. 6 (Ti) when 0.3 to 0.4 is
expected from analyses of the simpler spectra of
warmer subdwarfs 
(McWilliam 1997\markcite{McW97}).
 Given the crowded spectrum, the difference of
up to 0.3 dex from expectation is plausibly attributable to
measurement errors. The vanadium abundance [V/Fe] = 0.5
may be a reflection of our neglect of hyperfine splitting. 
There is marginal evidence for a mild enrichment of heavy
elements with observed and expected initial
abundances as follows:  [Y/Fe] = 0.2 and -0.2 expected,  [Zr/Fe] = 0.8
and 0.0, and [Ce/Fe] = 0.3 and -0.1.  

\subsection{WY\,Andromedae}

Photometry of WY\,And is reviewed by Zsoldos (1990\markcite{Zso90})
 who showed
the period  to be about 108 days. Rosino (1951\markcite{Ros51})
 and Joy (1952\markcite{Joy52})
found the spectral type to vary from G2 to K2. Our
spectrum shows TiO bands with a strength greater than in the spectrum
of UW\,Lib. Joy's radial
velocity of -191 km s$^{-1}$ is confirmed by our measurement
of -193 $\pm$ 1.1 km s$^{-1}$. Results of our abundance
analysis are summarized in Table 3.

Although there are no large differences between the observed
composition and that expected for a red giant with [Fe/H] = -1.0
 that has negotiated
the first dredge-up but not yet encountered the third dredge-up, 
the overall agreement with expectation is noticeably inferior
to that found for VW\,Eri and UW\,Lib.
 The $\alpha$-elements do not give the expected
uniform enhancement of about 0.3 dex: [Mg/Fe] = 0.1, 
[Si/Fe] = 0, [Ca/Fe] = -0.3, but [Ti/Fe] = 0.3. Aluminum and Mn
deficiencies are  in excess of expectation by about
0.2 dex: [Al/Fe]
= -0.3 and [Mn/Fe] = -0.6. Heavy elements with a dominant
contribution from the $s$-process are in the mean
enriched with [$s$/Fe] = 0.3 but two of the three elements with
5 or more lines show no enrichment. Europium is even underabundant
relative to iron: [Eu/Fe] = -0.4 from a single Eu\,{\sc ii}
line but [Eu/Fe] = 0.3 is expected. 
 We attribute these discrepancies to the
greater strength of the TiO bands in this more metal-rich
star, and to the enhanced probability of unsuspected blending
with TiO lines. It seems unlikely that the apparent abundance anomalies
in this or UW\,Lib are due to the dust-gas separation that greatly
affects some RV\,Tauri variables; Ca may be underabundant but Sc is not. 
Possibly, the star belongs to a stellar population whose initial composition
differs from the standard or expected composition. Certainly, subdwarfs with
[Fe/H] $\sim -1$ and anomalies relative to the standard composition
are now known (Nissen \& Schuster 1997\markcite{Nis97};
 Jehin et al. 1999\markcite{Jeh99}) but these
anomalies do not match well those found for WY\,And. Attribution of the
latter to errors of measurement seems the most likely explanation. 

Oxygen from the [O\,{\sc i}] 6300\AA\ and 6363\AA\ corresponds
to [O/Fe] = 0.7, a value that is approximately consistent with 
the latest estimates of the O abundance for metal-poor stars
(Israelian et al. 1998\markcite{Isr98};
 Boesgaard et al. 1999\markcite{Boe99}). A collection
of 3 C\,{\sc i} lines gives [C/Fe] = 0.3, which is the value
found by Gustafsson et al. (1999\markcite{Gus99})
 for disk dwarfs at [Fe/H] = -1.
This would suggest that there has been some C enrichment following
the first dredge-up's reduction of carbon on the first ascent of
the red giant branch.

\section{Concluding Remarks}

Our analyses of WY\,And, VW\,Eri, and UT\,Lib and our earlier
work on AB\,Leo, and SV\,Uma show that the subset of SRd variables
defined by weak metal lines and a high radial velocity 
are metal-poor supergiants with considerable similarities in
composition that are traceable to the corresponding similarity
of composition among metal-poor  dwarfs, subgiants, and giants
on the first red giant branch.  To our collection may be
added TY\,Vir (Luck \& Bond 1985\markcite{Luc85}) and CK\,Vir 
(Leep \& Wallerstein 1981\markcite{Lee81}).

In seeking the origins of the SRd variables, two obvious questions
arise: How are the stars related to red giants on the red giant
branch (RGB) and asymptotic giant branches (AGB)? What is the relationship
between the SRd variables and the RV\,Tauri variables?

Clues to the answers may be sought from the theoretical `HR diagram'
of log $g$ versus log T$_{\rm eff}$. Figure\,1 places our SRd stars in this 
diagram with RV\,Tauri stars drawn from our papers on the compositions
of these stars 
 (Giridhar, Rao
\& Lambert 1994\markcite{Gir94};
 Gonzalez, Giridhar \& Lambert 1997a\markcite{Gon97a}, 1997b\markcite{Gon97b};
Giridhar, Lambert \& Gonzalez 1998b\markcite{Gir98b},
 1999\markcite{Gir99}).
Since our analyses of RV\,Tauri and SRd variables have used common procedures,
systematic differences between the two kinds of stars are unlikely to be due
to errors in the analyses. Lines corresponding to constant luminosity
$L$ are drawn for log$L/L_{\odot}$ = 3.3 and 4.3 and a stellar
mass of 0.8$M_{\odot}$. A theoretical isochrone
for $Z$ = 0.0004 or [Fe/H] $\simeq -1.6$ is also shown (Bertelli et al. 1994
\markcite{Ber94}). 
 Isochrones for higher metallicity are displaced to lower
temperatures with very little change in the luminosity of the
most luminous stars on the RGB and AGB.
An increase from $Z$ = 0.0004 to 0.004 shifts the RGB tip from
(log$g$, logT$_{\rm eff}$) = (0.64, 3.64) to (0.22, 3.54), and the
AGB tip from (-0.1, 3.60) to (-0.4, 3.50). 
The $Z$ = 0.0004 isochrone is appropriate for VW\,Eri and AB\,Leo. The
other stars are more metal-rich with  the most metal-rich
(WY\,And) falling almost midway between $Z$= 0.0004 and 0.004.

Figure\,1 shows that the SRds are either  at the tip of
the RGB or  on the AGB at luminosities greater than the RGB tip.
Lloyd Evans (1975\markcite{Llo75}) noted that red variables
in metal-poor globular clusters are at the tip of a cluster's
giant branch.  
An AGB star may be distinguished by a C
and $s$-process enrichment provided that the third dredge-up 
has been activated. Our analysis is most accurate for VW\,Eri
which is neither C nor $s$-process enriched. There is a suspicion that
WY\,And may be enriched, as may be AB\,Leo and SV\,UMa from
Paper I. It remains to be discovered what sets an SRd variable apart
from essentially identical giants that are not variable. Perhaps,
they are stars that have begun a process of strong mass loss, or
have lower mass envelopes as a result of earlier history. 

The period ranges for SRd and RV\,Tauri stars are similar, and
the low temperature end of the RV\,Tauri sequence abuts the SRd
domain. It is difficult to accept that the stars are unrelated.
Observed abundance anomalies of some RV\,Tauri stars (see our papers cited
above) attributed to acquisition of dust-free gas by the stars  are not
in conflict with the suggested relation because our work (Giridhar, Lambert \&
Gonzalez 1999\markcite{Gir99}) shows that the anomalies are not found
in those RV\,Tauri stars that would be the closest relatives of SRd variables.
RV\,Tauri stars that are intrinsically metal-poor or cool do not exhibit
abundance anomalies.
 A relation is suggested by  Figure\,1.
RV\,Tauri variables from our series of papers on the
compositions of these stars are represented by different symbols for
the spectroscopic classes RV\,A, RV\,B, and RV\,C.
 It is
seen that the SRd's are at the low temperature boundary of the
RV\,Tauri regime which corresponds to a belt
bounded by lines of constant 
luminosity   for 
log$L/L_{\odot}$ = 3.3 and 4.3 assuming a stellar mass $M = 0.8M_{\odot}$.
 Evolution at constant luminosity
from the red giant branch is a signature of post-AGB evolutionary
tracks. The lower luminosity bound would correspond to evolution
from the tip of the first giant branch. The upper luminosity
bound suggests evolution off  the AGB. 
An interpretation of Figure\,1  is that SRds evolve into RV\,Tauri
variables as they cross the Hertzsprung gap from
low to high temperatures.
 Heavy mass loss has been
hypothesised as the cause of the early departure from the
AGB. Stellar pulsations
occurring in the SRd and RV\,Tauri variables may be the driver
of the mass loss.
Premature departure from the AGB 
is not an unexpected phenomenon
and has been invoked to account for the rarity of C-rich
AGB stars in the Magellanic Clouds.

This research has been supported in part by the Robert A.\, Welch Foundation of
Houston, Texas and the National Science Foundation (grant AST-9618414).

\clearpage
\clearpage
\section*{Figure Caption}
\figcaption{The log$g$ vs logT$_{\rm eff}$ diagram for SRd and RV\,Tauri
variables. The dashed lines are  tracks for a stellar mass of 0.8$M_{\odot}$
evolving at the constant luminosity of
log $L/L_{\odot}$ = 3.3 (lower
line) and 4.3 (upper line).
A theoretical isochrone for an age of 10$^{10}$ years and the
composition $Z$ = 0.0004 or [Fe/H] $\simeq$ -1.6 is shown with red giant, asymptotic giant,
and horizontal branches labeled.
 SRd and RV\,Tauri variables of spectroscopic classes
RV\,A, RV\,B, and RV\,C are distinguished by the symbols on the
legend on the figure.}
\end{document}